\documentclass[preprint,12pt]{elsarticle}

\usepackage{amssymb}
\usepackage{amsmath}
\usepackage{indentfirst}  %% 节标题后第一行段缩进
\usepackage{bm}    %%%%%  排印数学符号黑斜体时调用此宏包
\usepackage{graphicx}  %%%% 文中插入 eps 图形时调用此宏包，插图方法 1
\usepackage{cases}
\usepackage{extarrows}
\usepackage{booktabs}
\usepackage{tabularx}
\usepackage{threeparttable}
\usepackage{array}
\newcolumntype{C}[1]{>{\centering}p{#1}}
\begin{document}

\begin{frontmatter}



\title{One-way information reconciliation schemes of quantum key distribution}


\author{Li Yang}\ead{yang@is.ac.cn}
\author{Zhao Li}
\address{State Key Laboratory of Information Security, Institute of Information Engineering, Chinese Academy of Sciences, Beijing 100195, China}

\begin{abstract}
Information reconciliation(IR) is a basic step of quantum key distribution(QK\\ D). Classical message interaction is necessary in a practical IR scheme, and the communication complexity has become a bottleneck of QKD's development. Here we propose a concatenated method of IR scheme which requires only one time one-way communication to achieve any given error rate level. A QKD scheme with the concatenated IR can work without the special interactions of error rate estimation.
\end{abstract}


\end{frontmatter}

% \linenumbers

\section{Introduction}
\label{1}
After physical signal transmission, unconditionally secure key distribution protocol\cite{1,2} can be divided into three parts: advantage distillation\cite{3}, information reconciliation(IR)\cite{4} and privacy amplification\cite{5,6,7}. Quantum key distribution(QKD) is a mature unconditionally secure key distribution scheme with three phases: quantum signal transmission, raw key distillation(or advantage distillation), and classical data post-processing. IR is a basic step of classical data post-processing. Several IR protocols have been presented. In 1992, Bennett et al.\cite{8} proposed an IR protocol called Binary. Binary is simple and easy to operate, but it needs frequent interactive communication. It cannot find even errors in a block. In 1993, Brassard et al.\cite{9} proposed an IR protocol called Cascade, which can correct two errors in a block. Though its error correction ability is stronger than Binary, its computation and communication complexity is bigger. In 1999, Biham et al.\cite{10} proposed an IR scheme based on syndrome error correction. After that, Mayers et al.\cite{11} proposed an IR scheme based on error correcting code. Yang et al.\cite{12} suggested a key redistribution scheme for IR. These three IR protocols are non-interactive ones. In 2003, Buttler et al.\cite{17} proposed a IR scheme called Winnow. The number of the error correction rounds of Winnow is fewer than Binary and Cascade, but the error correction ability is limited.

It is clear that an IR needs to employ multi-round error correction to make the error rate arrive at an acceptable level in a practical QKD system. Since the problem an IR protocol deals with is not the errors of a bit string, but the bit inconsistence between two bit strings, we cannot use the well known concatenating error correction code directly. Binary, Cascade, and Winnow are all multi-rounds protocols. They adopt interactive communication to achieve an acceptable error rate level. However, the interactive communication causes extra time consuming, and becomes a bottleneck of the QKD's development. The non-interactive IR protocols such as that presented in \cite{10,11,12} are all one round error correction. They cannot achieve the practically acceptable low error rate. Thus it is necessary to construct new IR protocol. Here we propose a concatenating procedure for IR. The IR protocols designed based on this idea can reduce the error rate to any given level via only one time one-way communication, then they may improve the efficiency of a QKD's post-processing.

The techniques used in the construction of concatenated IR schemes are introduced in Sec.2. Some selection criterias of the error correction code in the concatenated method under a certain error rate of the channel is given in Sec.3. The construction method of a concatenated IR scheme with three examples is given in Sec.4. Some discussions and the conclusion are given in Sec.5 and Sec.6, respectively.

\section{Preliminaries}
\subsection{Wire link permutation}
Wire link permutation(WLP) is a simple and fast digital circuit bit-permutation technique, without the help of gate circuits. There are many different WLPs. We can see that, in an IR protocol, it is necessary to do a random bit-permutation between any two successive error correction rounds. The permutation used in an IR protocol should be as uniform as possible, that means the bits in a block should be dispersed uniformly into different blocks after a permutation. A proper WLP is shown in Fig. 1.

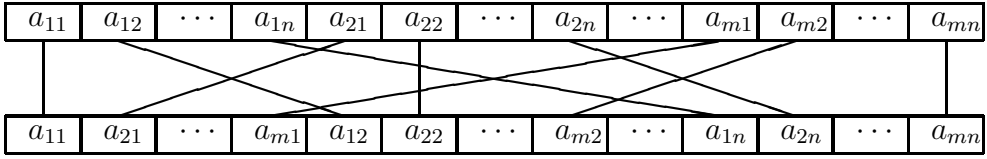
\begin{figure}[!htbp]
  \begin{center}
  \setlength{\unitlength}{1cm}
  \begin{picture}(13,2.5)
  \thicklines
  \multiput(0,0)(0,0.5){2}{\line(1,0){13}}
  \multiput(0,0)(1,0){14}{\line(0,1){0.5}}

   \put(0.3,0.2){$a_{11}$}      \put(7.3,0.2){$a_{m2}$}
   \put(1.3,0.2){$a_{21}$}      \put(8.3,0.2){$\cdots$}
   \put(2.3,0.2){$\cdots$}      \put(9.3,0.2){$a_{1n}$}
   \put(3.3,0.2){$a_{m1}$}      \put(10.3,0.2){$a_{2n}$}
   \put(4.3,0.2){$a_{12}$}      \put(11.3,0.2){$\cdots$}
   \put(5.3,0.2){$a_{22}$}      \put(12.3,0.2){$a_{mn}$}
   \put(6.3,0.2){$\cdots$}

  \multiput(0,1.5)(0,0.5){2}{\line(1,0){13}}
  \multiput(0,1.5)(1,0){14}{\line(0,1){0.5}}

   \put(0.3,1.7){$a_{11}$}      \put(7.3,1.7){$a_{2n}$}
   \put(1.3,1.7){$a_{12}$}      \put(8.3,1.7){$\cdots$}
   \put(2.3,1.7){$\cdots$}      \put(9.3,1.7){$a_{m1}$}
   \put(3.3,1.7){$a_{1n}$}      \put(10.3,1.7){$a_{m2}$}
   \put(4.3,1.7){$a_{21}$}      \put(11.3,1.7){$\cdots$}
   \put(5.3,1.7){$a_{22}$}      \put(12.3,1.7){$a_{mn}$}
   \put(6.3,1.7){$\cdots$}

   \put(0.5,0.5){\line(0,1){1}}
   \put(1.5,1.5){\line(3,-1){3}}
   \put(3.5,1.5){\line(6,-1){6}}
   \put(1.5,0.5){\line(3,1){3}}
   \put(5.5,0.5){\line(0,1){1}}
   \put(7.5,1.5){\line(3,-1){3}}
   \put(3.5,0.5){\line(6,1){6}}
   \put(7.5,0.5){\line(3,1){3}}
   \put(12.5,0.5){\line(0,1){1}}

 \end{picture}
 \end{center}
 \caption{\label{lz}The wire link permutation $W$ adopted in our scheme.}
\end{figure}

We can see that after the permutation $W$ the first bit of the first block $(a_{11}, a_{12}, ..., a_{1n})$ is put in the first position in the new round; The first bit of the second block $(a_{21}, a_{22}, ..., a_{2n})$ is put in the second position in the new round, etc.; Go on like this until the last block $(a_{m1}, a_{m2}, ..., a_{mn})$: the first bit $a_{m1}$ is put in the $m^{th}$ position in the new round, etc..

The WLP should be done between each pair of successive error correction rounds. The $i^{th}$ permutation $W^{i}$ is as follows,

\begin{equation}
\begin{array}{r@{~}l}
& (a_{11}^{(i)}, a_{12}^{(i)}, \cdots , a_{1n}^{(i)}, a_{21}^{(i)}, a_{22}^{(i)}, \cdots , a_{2n}^{(i)}, \cdots\cdots , a_{m1}^{(i)}, a_{m2}^{(i)}, \cdots , a_{mn}^{(i)}) \\
\xlongrightarrow{\tiny{W^{(i)}}}
 & (a_{11}^{(i)}, a_{21}^{(i)}, \cdots , a_{m1}^{(i)}, a_{12}^{(i)}, a_{22}^{(i)}, \cdots , a_{m2}^{(i)}, \cdots\cdots , a_{1n}^{(i)}, a_{2n}^{(i)}, \cdots , a_{mn}^{(i)}).
\end{array}
\end{equation}

We can rearrange the data string $(a_{11}^{(i)}, a_{12}^{(i)}, \cdots , a_{1n}^{(i)}, a_{21}^{(i)}, a_{22}^{(i)}, \cdots , a_{2n}^{(i)}, \cdots\\\cdots , a_{m1}^{(i)}, a_{m2}^{(i)}, \cdots , a_{mn}^{(i)})$ into a matrix as

\begin{equation*}
A^{(i)}\triangleq
\begin{bmatrix}
  a_{11}^{(i)} & a_{12}^{(i)} & \cdots & a_{1n}^{(i)} \\
  a_{21}^{(i)} & a_{22}^{(i)} & \cdots & a_{2n}^{(i)} \\
  \hdotsfor{4}\\
  a_{m1}^{(i)} & a_{m2}^{(i)} & \cdots & a_{mn}^{(i)}
\end{bmatrix}
.
\end{equation*}
It can be seen that every row is a codeword before the permutation, and every column is a codeword after the permutation. Since the $W^{(i)}$ changes the rows to the columns, it is just a transpose operation of the matrix $A^{(i)}$. Thus,
\begin{equation*}
W^{(1)}=\cdots=W^{(i)}=\cdots\triangleq W,
\end{equation*}
\begin{equation*}
W^{-1}=W.
\end{equation*}

\subsection{Non-interactive IR schemes}
There are three kinds of non-interactive IR schemes. The first one is the syndrome IR scheme\cite{10}. In this scheme, Alice sends syndromes to do error correction. Bob uses the equation $s_A\oplus s_B=H(K_A\oplus K_B)$ to correct his raw key $K_B$ to Alice's raw key $K_A$. The second one is the IR scheme of Mayers\cite{11}. In this scheme, Alice encodes a local random string $x$ to get the codeword $c$, and uses her raw key $K_A$ to do one time pad with it to get $c\oplus K_A$. Then she sends it to Bob. Bob adds his raw key $K_B$ to it to get the $(c\oplus K_A)\oplus K_B=c\oplus e$, and decodes it to get the codeword $c$. Then he adds it to the receiving $c\oplus K_A$ to get $K_A$. The third one is the key redistribution scheme\cite{12}. The basic idea of this scheme is: Alice first encodes a local random bit string with an error correcting code, then she uses her raw key to do one time pad with the codeword and transmits it to Bob. Bob adds his raw key to the received bit string and decodes the error correcting code to get Alice's local random bit string, then takes it as the secret key between them. The whole protocol can be summarized as follows.

\begin{enumerate}
\item Alice generates a random bit string $x$.
\item Alice uses a generator matrix $g$ to encode $x$ and gets the code word $c$, where $g$ is a globe public parameter.
\item Alice uses the raw key $K_a$ to do bitwise XOR operation with the code string $c$ to get $K_a\oplus c$. Then she transmits it to Bob.
\item Bob does the same operation to the received string with $K_b$ and gets $(c\oplus K_a)\oplus K_b=c\oplus e$. He uses check matrix $h$ and $c\oplus e$ to calculate the syndrome $s$. Using $s$, he gets the error vector $e$ and the codeword $c$. Then he gets the random bit string $x$ by decoding $c$, and takes it as the secret key between them.
\end{enumerate}

If the generator matrix is kept secret, the key redistribution protocol may generate a secure final key. It can also realize group oriented key distribution, personal identification, and message authentication for non-broadcast channel via key-controlled error-correcting code. Thus the key redistribution protocol may realize the IR and the privacy amplification in one step.

\subsection{Classical message authentication using CRC-based MAC\cite{13,14}}
CRC-based MAC designed for stream cipher is a scheme with information-theoretic security based on cyclic redundancy code(CRC). LFSR can be used to realize rapid polynomial division in a CRC authentication scheme. This kind of authentication scheme can authenticate large amount of messages by consuming a few bits of the key. For this reason, we suggest using it to authenticate the classical channel of QKD. The CRC based authentication scheme is as follows.

Denote the $n$ bits message to be authenticated as $M$. Make $M=M_{n-1}...M_{1}M_{0}$ and the polynomial $M(x)=\sum_{i=0}^{n-1}M_ix^i$ associated. Denote the CRC hash function as $h$, and the MAC value as $aut$. The output of $h$ is an $m$ bit string.

\begin{enumerate}
\item Alice and Bob secretly preshare a binary irreducible polynomial $p(x)$ of degree $m$, and a $m$-bit random string $K$ as their one time pad key.
\item Alice calculates $h(M)=coef(M(x)\cdot x^m\, mod\: p(x))$.
\item Alice gets the $m$-bit $aut$ of $M$ by calculating $h(M)\oplus K$.
\item Alice sends $aut$ and $M$ to Bob
\item Bob uses the received $M'$ to calculate a $aut''$, and checks whether it is equal to the $aut'$ he received.
\end{enumerate}

The successful attack probability is $\frac{n+m}{2^{m-1}}$\cite{13} for any $n$ and $m>1$.

\subsection{Hamming code\cite{15}}
$[n,n-k,3]$Hamming code  over $F_2$ with $n=2^k-1$ has fast error correction algorithm for its special structure. Given a serial number from 1 to $n$ to denote the position of each bit in a codeword. The check bits are inserted into $2^lth$ positions, where $0\leq l< k$. The left positions are information bits. Its generator matrix is obtained by exchanging the $2^lth$ column with the $(n-l)th$ column of the corresponding systematic code respectively, where $0\leq l< k$. The decoding method is multiplying the receiving bit-string with the parity check matrix to get the syndrome $s=(s_1,...,s_k)$, then the binary number $(s_1...s_k)_2$ indicates just the position of an error bit in the codeword.

Consider of the fast decoding algorithm of Hamming code, we choose it as the error-correcting code to be concatenated in our concatenated IR scheme.

\section{Some selection criteria of concatenated IR schemes}
Usually, after one error correction round, we can hardly reduce the error rate to an acceptable level, thus we have to do more error correction rounds. Binary, Cascade and Winnow include multi-round error correction, and need a parity check before every round to determine whether a block needs to be corrected. The necessary interactive communication makes the efficiency of these protocols decreased. The original scheme of Biham\cite{10}, Mayers\cite{11} and key redistribution\cite{12} employ only one-round error correction, which cannot reduce the error rate to an acceptable level in practical system. In order to realize both one time one-way communication and an acceptable error rate level simultaneously, we suggest a concatenating method of IR. All the three one round IR protocols can be reconstructed based on this idea. In this section, we will prove some selection criteria for choosing the number of round and the error correcting code under a given error rate of the channel.

\textbf{Definition 1}\cite{16}. Let $C$ be a linear code of length $n$ and let $A_i$ be the number of codewords of weight $i$, then
\begin{equation}
    A(z,n):= \sum_{i=0}^{n}A_iz^i
\end{equation}
is called the weight enumerator of $C$. The sequence $(A_i)^{n}_{i=0}$ is called the weight distribution of $C$. If $C$ is linear and $\vec{c}\in C$, then the number of codewords at distance $i$ from $\vec{c}$ equals $A_i$.

For binary Hamming code of length $n$, the weight enumerator
\begin{equation}
    A(z,n)=\sum^{n}_{i=0}A_iz^i=\frac{1}{n+1}(1+z)^n+\frac{n}{n+1}(1+z)^{\frac{n-1}{2}}(1-z)^{\frac{n+1}{2}}.
\end{equation}
It should be noticed that, for Hamming code, $n=2^k-1$ is an odd number.
From Eq.(3), compare the polynomial coefficients of the two sides of Eq.(3), we get that $A_1=A_2=A_{n-2}=A_{n-1}=0$, and all other coefficients are non-zero integers. For example, for the code $[7,4,3], n=7$, we get $A(z,7)=1+7z^{3}+7z^{4}+z^{7}$. For the code $[15,11,3], n=15$, we get $A(z,15)=1+35z^3+105z^4+168z^5+280z^6+435z^7+z^{15}+35z^{12}+105z^{11}+168z^{10}+280z^{19}+435z^{8}$.

According to Eq.(2), we calculate the weight distribution $(A_i)^n_{i=0}$ of Hamming code of length $n$.
\begin{eqnarray}
A(z,n) & = & \frac{1}{n+1}(1+z)^n+\frac{n}{n+1}(1+z)^{\frac{n-1}{2}}(1-z)^{\frac{n+1}{2}}\nonumber\\
           & = & \frac{1}{n+1}\sum^{n}_{k=0}C_n^kz^k+\frac{n}{n+1}(1-z)\sum^{\frac{n-1}{2}}_{i=0}C^i_{\frac{n-1}{2}}(-1)^iz^{2i}\nonumber\\
           & = & \frac{1}{n+1}\sum^{n}_{k=0}C_n^kz^k+\frac{n}{n+1}\sum^{\frac{n-1}{2}}_{i=0}[(-1)^iC^i_{\frac{n-1}{2}}z^{2i}+(-1)^{i+1}C^i_{\frac{n-1}{2}}z^{2i+1}]\nonumber\\
           & = &
           \frac{1}{n+1}\sum^{n}_{k=0}C_n^kz^k+\frac{n}{n+1}\sum^{n}_{k=0}(-1)^{\lceil \frac{k}{2}\rceil}C^{\lfloor \frac{k}{2}\rfloor}_{\frac{n-1}{2}}z^{k}\nonumber\\
           & = &
           \sum^{n}_{k=0}(\frac{1}{n+1}C_n^k+\frac{n}{n+1}(-1)^{\lceil \frac{k}{2}\rceil}C^{\lfloor \frac{k}{2}\rfloor}_{\frac{n-1}{2}})z^{k}
\end{eqnarray}
Compare the coefficients with $A(z,n)=\sum^n_{k=0}A_kz^k$, we get$$ A_k=\frac{1}{n+1}C_n^k+\frac{n}{n+1}(-1)^{\lceil \frac{k}{2}\rceil}C^{\lfloor \frac{k}{2}\rfloor}_{\frac{n-1}{2}}.$$

\textbf{Definition 2}\cite{16}. Let $C\subseteq Q^n$ be a code with $M$ words. We define
\begin{equation}
    A_i:=M^{-1}|\{(\vec{x},\vec{y})|\vec{x}\in C, \vec{y}\in C, d(\vec{x},\vec{y})=i\}|.
\end{equation}
The sequence $(A_i)^n_{i=0}$ is called the distance distribution or inner distribution of $C$.

Note that if $C$ is linear, the distance distribution is weight distribution. Thus, for Hamming code, the weight distance and the distance distribution are the same. With the weight distribution of Hamming code calculated in Eq.(2), we get that its distance distribution is $(A_k)^n_{k=0}$, here $A_k=\frac{1}{n+1}C_n^k+\frac{n}{n+1}(-1)^{\lceil \frac{k}{2}\rceil}C^{\lfloor \frac{k}{2}\rfloor}_{\frac{n-1}{2}}, k=0, 1, \cdots, n$. This means, for any Hamming code $\vec{c}$ of length $n$, the number of the codewords at distance $i$ from $\vec{c}$ is $A_i, i=0, 1, \cdots, n$.

Suppose using Hamming code of length $n$, bit error probability is $p$, the expected number of errors per block before decoding is $np$.

(a)If one error occurs, the number of error bits is zero after error correction.

(b)If $k, (2\leq k\leq n-1)$ errors occur, there are two situations when executing error correction:

\begin{enumerate}
  \item The $k$ errors turn one codeword into another codeword. In this situation, we cannot use error-correcting code to correct any bit of errors. There are still $k$ errors after error correction. For any Hamming codeword $\vec{c}$ of length $n$, the number of the codewords at distance $k$ from $\vec{c}$ is $A_k$. Thus, the probability of this situation is $A_kp^k(1-p)^{n-k}$. This means there will be still $k$ errors with probability $A_kp^k(1-p)^{n-k}$ after error correction.
  \item The $k$ errors do not turn the code into another code. In this situation, the error correction may correct only one error to reduce the number of error to $k-1$. But also, this may cause a new error to increase the number of error to $k+1$. This means we can get a new codeword at distance $k-1$ from codeword $\vec{c}$ or a new codeword at distance $k+1$ from codeword $\vec{c}$. For any codeword $\vec{c}$, the number of the codewords whose distance with $\vec{c}$ is $k-1$ or $k+1$ are separately $A_{k-1}, A_{k+1}$. Thus, after error correction we can get one of the $A_{k-1}+A_{k+1}$ codewords. Suppose each codeword can be gotten with the same probability in the error correction. After error correction the probability of reducing the error number to $k-1$ is $\frac{A_{k-1}}{A_{k-1}+A_{k+1}}$, and the probability of increasing the error number to $k+1$ is $\frac{A_{k+1}}{A_{k-1}+A_{k+1}}$. The probability that $k$ errors do not turn the codeword $\vec{c}$ to another codeword is $(C^k_n-A_k)p^k(1-p)^{n-k}$ because the number of the codewords at distance $k$ from $\vec{c}$ is $A_k$. Thus, the probability that $k$ errors cannot turn a codeword to another codeword and the number of errors is reduced to $k-1$ is $(C^k_n-A_k)\frac{A_{k-1}}{A_{k-1}+A_{k+1}}p^k(1-p)^{n-k}$. The probability that $k$ errors cannot turn a codeword to another codeword and the number of errors is increased to $k-1$ is $(C^k_n-A_k)\frac{A_{k+1}}{A_{k-1}+A_{k+1}}p^k(1-p)^{n-k}$.

      (c)When $n$ errors occur, for $A_n=1$, this means the number of the codewords at distance $n$ with $\vec{c}$ is $1$. The length of the codeword is $n$, thus if all of the $n$ bits are wrong, there is only $C^n_n=1$ situation. Thus $n$ errors can only turn a codeword to another codeword. In this situation after error correction there are still $n$ errors. The probability of this situation is $p^n$.
\end{enumerate}

From the above analysis, we can calculate the mathematical expectation of the errors in each block after error correction. Let the bit error probability is $p_1$ after error correction. Thus after error correction the mathematical expectation of errors in each block is $np_1$.
\begin{eqnarray}
np_1 & = & \sum_{k=2}^{n-1}[kA_kp^k(1-p)^{n-k}+(k-1)(C^k_n-A_k)\frac{A_{k-1}}{A_{k-1}+A_{k+1}}p^k(1-p)^{n-k}+\nonumber\\
     &   &
     (k+1)(C^k_n-A_k)\frac{A_{k+1}}{A_{k-1}+A_{k+1}}p^k(1-p)^{n-k}]+np^n\nonumber\\
     & = & \sum_{k=2}^{n-1}[kA_k+(C^k_n-A_k)\frac{(k-1)A_{k-1}+(k+1)A_{k+1}}{A_{k-1}+A_{k+1}}]p^k(1-p)^{n-k}+nA_np^n\nonumber\\
     & = &
     \sum_{k=0}^{n}[kA_k+(C^k_n-A_k)(k+\frac{A_{k+1}-A_{k-1}}{A_{k-1}+A_{k+1}})]p^k(1-p)^{n-k}.
\end{eqnarray}
Here, denote $A_{-1}=0, A_{n+1}=0$. When $A_{k+1}=A_{k-1}=0$, denote $\frac{A_{k+1}-A_{k-1}}{A_{k-1}+A_{k+1}}=0$.

From the above equation, we can get
\begin{eqnarray}
np_1 & = &
     \sum_{k=0}^{n}[(C^k_n-A_k)\frac{A_{k+1}-A_{k-1}}{A_{k-1}+A_{k+1}}+kC^k_n]p^k(1-p)^{n-k}\nonumber\\
     & = &
     \sum_{k=0}^{n}(C^k_n-A_k)\frac{A_{k+1}-A_{k-1}}{A_{k-1}+A_{k+1}}p^k(1-p)^{n-k}+np.
\end{eqnarray}
Thus, $p_1<p$ equals the following equation
\begin{equation}
    \sum_{k=0}^{n}(C^k_n-A_k)\frac{A_{k+1}-A_{k-1}}{A_{k-1}+A_{k+1}}p^k(1-p)^{n-k}<0.
\end{equation}

For the Hamming code of length $n=7$, we have
\begin{equation}
    7p_1=63p^2-182p^3+210p^4-84p^5.\nonumber
\end{equation}
\begin{equation}
    p_1=9p^2-26p^3+30p^4-12p^5.
\end{equation}
From $p_1<p$, we get
\begin{equation}
    0<p<\frac{1}{6}(3-\sqrt{3}), \textrm{or}\, \frac{1}{2}<p<\frac{1}{6}(3-\sqrt{3}).
\end{equation}

This means we can use error-correcting code to reduce the error rate if and only if the bit error probability $p$ satisfies $0<p<\frac{1}{6}(3-\sqrt{3})$ or $\frac{1}{2}<p<\frac{1}{6}(3-\sqrt{3})$.
\vspace*{2mm}

\centerline{\includegraphics{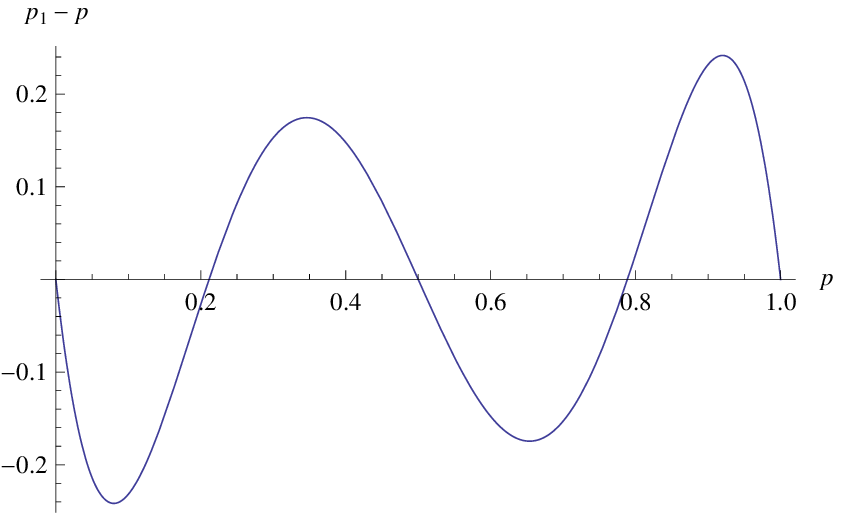}}

\begin{center}
\parbox{15.5cm}{\small{ Figure 2:}  The error rate after error-correction $p_1$ varies with the inial error rate $p$ \\when $n=7$. }
\end{center}

From Fig.2 we can see there are five points of intersection between the curve and X-axis. They are $0, \frac{1}{6}(3-\sqrt{3}), \frac{1}{2}, \frac{1}{6}(3-\sqrt{3}), 1$. If the $p$ is in the interval $[\frac{1}{6}(3-\sqrt{3}), \frac{1}{2}, \frac{1}{6}(3-\sqrt{3})]$, $p_1$ will go forwards to $\frac{1}{2}$ after error correction. In this situation we cannot correct the errors. The interval of $p$ where we can use this code is $[0, \frac{1}{6}(3-\sqrt{3})]$ and $[\frac{1}{6}(3-\sqrt{3}), 1]$.

The error rate after error-correction $p_1$ varying with the inial error rate $p$\\ when $n=15$ is as Fig.2.
\vspace*{2mm}

\centerline{\includegraphics{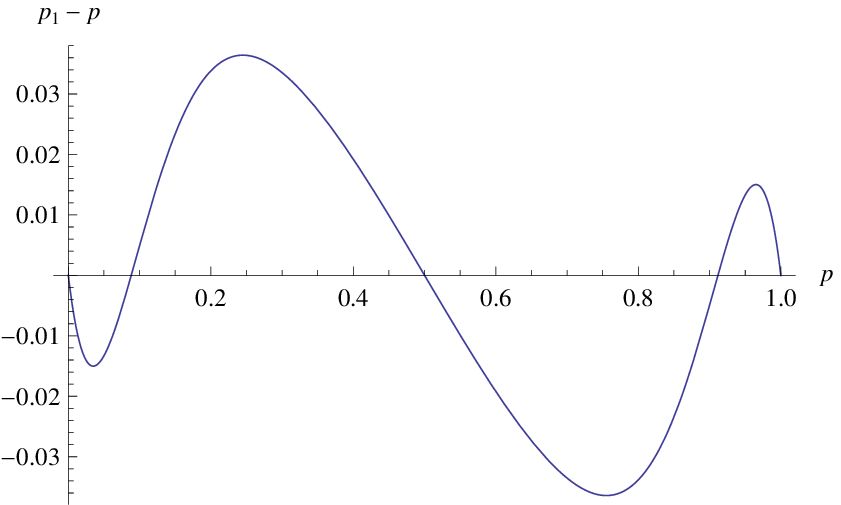}}

\begin{center}
\parbox{15.5cm}{\small{Figure 3:}  The error rate after error-correction $p_1$ varies with the inial error rate $p$ \\when $n=15$. }
\end{center}

Compare Fig.3 with Fig.2 we can see the effective interval of Hamming code $[15, 11, 3]$ is less than that of Hamming code $[7, 4, 3]$.

\textbf{Lemma 1.} Let $C$ be the $[n,n-k,3]$ Hamming code over $F_2$, where $n=2^k-1$. Suppose the upper bound of the average number of errors within per block after one error correction round with $C$ is $\chi$, then we have
\begin{equation}
\chi = 1+np-2p^n-(1-p+2np)(1-p)^{n-1},
\end{equation}
where $p$ is the bit error rate of the channel.

\textbf{Lemma 2.} $\chi < n(n-1)p^2[1+\frac{1}{2}(1-p)^{n-2}].$

\textbf{Theorem 1.} When $C$ is used as the error correcting code, if bit error rate $p$ satisfies the condition $p<\frac{1}{(n-1)[1+\frac{1}{2}(1-p)^{n-2}]}$, then the concatenated error correction scheme can achieve any given error rate level.

\textbf{Corollary 1.} If bit error rate $p<p_{th}=\frac{2}{3(n-1)}$, the concatenated error correction scheme can reduce the error rate to any given level.

Table 1 and 2 show the concatenating results based on Eq.(8), which are useful for choosing the proper error correcting code and the concatenating depth $l$. Parameter $\eta$ is the information rate of the concatenated IR algorithm. $\alpha$ is the final error rate of the concatenated IR algorithm. It is required that after $l$ rounds error correction the final error rate $\alpha$ should be below $1\times 10^{-9}$. According to this criterion, the required error correction round $l$ and the final left bit rate are determined. The results based on Hamming code $[15,11,3]$ and $[7,4,3]$ are given in Table 1 and Table 2, respectively.

\begin{table}[!htbp]
%\begin{threeparttable}[b]
\tiny
\caption{Concatenated IR based on $[15,11,3]$ code. $p$ represents the channel error rate. $l$ represents the needed error correction rounds. $\alpha$ represents the final error rate. $\eta$ represents the left bit rate.}
\vspace{-2mm}
\center
\arrayrulewidth=1pt
\begin{tabular}{C{0.06cm}C{1.3cm}C{1.3cm}C{1.3cm}C{1.3cm}C{1.3cm}C{1.3cm}C{1.3cm}}\toprule
%\hline
%\hline
\addlinespace[5pt]
$p$ & 0.01 & 0.02 & 0.04 & 0.05 & 0.06 & 0.07 & 0.08\tabularnewline
%\cmidrule{1-7}\morecmidrules\cmidrule{1-7}
\midrule
$l$ & 4 & 5 & 6 & 7 & 8 & 9 & 11 \tabularnewline
\addlinespace[4pt]
$\eta$ & 0.289 & 0.212 & 0.156 & 0.114 & 0.084 & 0.061 & 0.024 \tabularnewline
%\colrule
%\hline
\addlinespace[4pt]
$\alpha$ & $3.58\!\times\! 10^{-13}$ & $2.59\!\times\! 10^{-15}$ & $1.77\!\times\! 10^{-12}$ & $2.72\!\times\! 10^{-14}$ & $8.86\!\times\! 10^{-15}$  & $2.35\!\times\! 10^{-11}$ & $2.04\!\times\! 10^{-11}$  \tabularnewline\bottomrule
%\hline
%\hline
\end{tabular}
%\begin{tablenotes}
%\item
%\end{tablenotes}
%\end{threeparttable}
\end{table}

\begin{table}[!htbp]
%\begin{threeparttable}[b]
\tiny
\caption{Concatenated IR based on $[7,4,3]$ code. $p$ represents the channel error rate. $l$ represents the needed error correction rounds. $\alpha$ represents the final error rate. $\eta$ represents the left bit rate.}
\vspace{-2mm}
\center
\arrayrulewidth=1pt
\begin{tabular}{C{0.06cm}C{1.3cm}C{1.3cm}C{1.3cm}C{1.3cm}C{1.3cm}C{1.3cm}C{1.3cm}C{1.3cm}C{1.3cm}C{1.3cm}}\toprule
%\hline
%\hline
\addlinespace[5pt]
$p$ & 0.05  & 0.07 & 0.09 & 0.10 & 0.12 & 0.13 & 0.14\tabularnewline
%\cmidrule{1-7}\morecmidrules\cmidrule{1-7}
\midrule
$l$ & 5 & 5 & 6 & 6 & 7 & 7 & 8 \tabularnewline
\addlinespace[4pt]
$\eta$ & 0.061 & 0.061 & 0.035 & 0.035 & 0.020 & 0.011 & 0.011 \tabularnewline
%\colrule
%\hline
\addlinespace[4pt]
$\alpha$ & $5.22\!\times\! 10^{-14}$ & $5.93\!\times\! 10^{-10}$ & $1.20\!\times\! 10^{-12}$ & $1.74\!\times\! 10^{-10}$ & $1.66\!\times\! 10^{-12}$ & $6.96\!\times\! 10^{-10}$ & $1.04\!\times\! 10^{-13}$ \tabularnewline\bottomrule
%\hline
%\hline
\end{tabular}
%\begin{tablenotes}
%\item
%\end{tablenotes}
%\end{threeparttable}
\end{table}

If the channel error rate $p$, the final error rate $\alpha$ and the error correcting code are given, the concatenating depth $l$ will be determined.

\section{The construction of concatenated IR schemes}
Based on the selection criteria given in Sec. 3, three IR scheme\cite{10,11,12} are constructed with the concatenating method as follows.

I.  Firstly we consider the reconstruction of Biham's syndrome error correction protocol\cite{10}. Follow Winnow\cite{17}, we choose $[n,n-k,3]$ Hamming code. Currently a typical error rate for a QKD IR protocols to deal with is less than 5\%. According to Theorem 1, we can choose [15, 11, 3] Hamming code as the basic code, whose error correction ability is 6.7\%. The protocol is as follows.

\begin{enumerate}
\item Alice divides the raw key string into 15-bit length blocks and then performs the permutation $W$ on it. Alice calculates the syndromes $s_{Ai}^{(j)}$, and discards the check bits of each block, here $i$ is the serial number of the block, and $j$ is the serial number of the round. Alice repeats above operations from $j=1$ to $j=l$, to get the syndromes $s_{Ai}^{(1)}, ..., s_{Ai}^{(l)}, i=1,\cdots , n$, where $l$ is the predetermined number of the correction rounds. The Alice's final bit-string is the common random string to be privacy amplified.
\item Alice takes the syndromes $s_{Ai}^{(1)},s_{Ai}^{(2)}, ..., s_{Ai}^{(l)}$ $(i=1,\cdots ,n)$ as her message to be sent. She uses CRC authentication algorithm to calculate the MAC of the message and sends the MAC and the message to Bob.
\item After receiving the sequence $s_{Ai}^{(1)},s_{Ai}^{(2)}, ..., s_{Ai}^{(l)}$, Bob uses the CRC authentication algorithm and the one time pad key $K$ to check whether the message is coming from Alice and has not been changed. If the authentication is passed, Bob uses the wire link permutation $W$ to transform his raw key and calculates the syndrome $s_{Bi}^{(1)}$ of every block. Then he calculates the $i^{th}$ syndrome $s_{i}^{(1)}=s_{Ai}^{(1)}\oplus s_{Bi}^{(1)}$, and does error correction to the $i^{th}$ block, $i=1,\cdots , n$. After the error correction of the first round he discards all the check bits. Bob repeats above operation to get the syndromes $s_{i}^{(j)}, i=1,\cdots , n$ and performs error correction from $j=1$ to $j=l$. Finally he gets the key of Alice after $l$ rounds error correction.
\end{enumerate}

Suppose the initial error rate is 3\%. According to the criteria in Sec. 3, we get the upper bound of the final error rate and the final bit rate after each error correction round, as shown in Table 3.

\begin{table}[!htbp]
%\begin{threeparttable}[b]
\scriptsize
\caption{The upper bound of error rate based on Lemma 1 and the left bit rate after each error correction round. Suppose channel error rate is 3\%. The chosen code is Hamming code $[15,11,3]$. The data in this table are the upper bound of error rate and left bit rate after $i$ rounds error correction, $1\leq i\leq 6$.}
\vspace{-2mm}
\center
\arrayrulewidth=1pt
\begin{tabular}{ccccccc}\toprule
%\hline
%\hline
\addlinespace[5pt]
Round & 1 & 2 & 3 & 4 & 5 & 6 \tabularnewline
%\cmidrule{1-7}\morecmidrules\cmidrule{1-7}
\midrule
%\colrule
%\hline
Error Rate & $1.53\!\times\! 10^{-2}$ & $4.40\!\times\! 10^{-3}$ & $3.93\!\times\! 10^{-4}$ & $3.23\!\times\! 10^{-6}$ & $2.20\!\times\! 10^{-10}$ & $5.92\!\times\! 10^{-17}$ \tabularnewline
\addlinespace[4pt]
Left Rate & $0.733$ & $0.538$ & $0.394$ & $0.289$ & $0.212$ & $0.156$ \tabularnewline\bottomrule
%\hline
%\hline
\end{tabular}
%\begin{tablenotes}
%\item
%\end{tablenotes}
%\end{threeparttable}
\end{table}

The concatenating depth $l$ in the protocol is determined by a given final error rate. Table 3 shows that when the concatenating depth $l$ is 5, we can get an error rate under $1.0\times 10^{-9}$ with a left bit rate 0.212.

\vspace{3mm}
II.  The original key redistribution protocol can be reconstructed as follows.
\begin{enumerate}
\item Alice generates a random string $r_A^{(1)}$, and divides it into blocks with length 11, $r_{A}^{(1)}=(r_{1}^{(1)}, \cdots ,r_{n_{1}}^{(1)})$. She uses the [15, 11, 3] Hamming code to encode each block and gets $c^{(1)}=(c_{1}^{(1)}, \cdots ,c_{n_{1}}^{(1)})$, and then uses the wire link permutation $W$ to rearrange $c^{(1)}$. She divides it again into blocks with length 11, $r_{A}^{(2)}=(r_{1}^{(2)}, \cdots ,r_{n_{2}}^{(2)})$. Executing those operations $l$ rounds, she gets the codeword string $c^{(l)}=(c_{1}^{(l)}, \cdots ,c_{n_{l}}^{(l)})$. There is no permutation in the last round. The above process can be written as
    $$C_l[P_{l-1}[C_{l-1}\cdots [C_2[P_1[C_1(r_A^{(1)})]]]\cdots ]]=c^{(l)},$$
    where $P_i$ is the $i^{th}$ round wire link permutation $W$, $C_i$ is the $i^{th}$ round encoding with $[15,11,3]$ code.
\item Alice uses her raw key $K_{A}$ to xor bit by bit with the codeword string $c^{(l)}$, and gets $K_{A}\oplus c^{(l)}$. It is the message to be sent. She uses CRC authentication algorithm to calculate the MAC of the message, and sends the MAC and the message to Bob.
\item Bob uses the CRC authentication algorithm to check whether the message has been changed. If the authentication is passed, Bob uses his raw key $K_{B}$ to do xor bit by bit with the received codeword string and gets $(K_{A}\oplus c^{(l)})\oplus K_{B}=c^{(l)}\oplus e$. Bob decodes it and does the inverse wire link permutation $W^{-1}=W$. He repeats above operations round by round, and gets $r_{B}^{(1)}$ finally. Here we require $\frac{W_H(r_{B}^{(1)}\oplus r_{A}^{(1)})}{|r_{A}^{(1)}|}\leq 1.0\times 10^{-9}$.
\end{enumerate}

The concatenating depth $l$ is also 5 according to Table 3.

\vspace{3mm}
III.  The concatenated version of Mayer's ECC-based IR protocol is as follows.

1-3.\quad The same as that of the key redistribution protocol.

 4.\qquad \!Bob uses the $r_B^{(1)}$ to do concatenated encoding just as Alice has

\qquad\quad \!\!done to get
 $$c'^{(l)}=C_l[P_{l-1}[C_{l-1}\cdots [C_2[P_1[C_1(r_B^{(1)})]]]\cdots ]],$$

\qquad\quad \!\!and gets the $K_A'$ by calculating $(K_{A}\oplus c^{(l)})\oplus c'^{(l)}$. Here we require

\qquad\quad \!\!$\frac{W_H(K_A\oplus K_A')}{|K_A|}\leq 1.0\times 10^{-9}$, that means $\frac{W_H(c^{(l)}\oplus c'^{(l)})}{|c^{(l)}|}\leq 1.0\times 10^{-9}$.

The concatenating depth $l$ is also 5 according to Table 3. The step 4 shows that the concatenated ECC-based IR protocol needs to do an extra concatenated encoding. In step 3, Bob uses his raw key $K_{B}$ to do xor bit by bit with the received sequence and gets $(K_{A}\oplus c^{(l)})\oplus K_{B}=c^{(l)}\oplus e$. He gets gradually all the vectors $e^{(l)},e^{(l-1)}...,e^{(1)},c_B^{(l)},c_B^{(l-1)},...,c_B^{(1)},$ and $r_B^{(1)}$ in the end. His purpose is getting $K_A$, so he should get $e$ and then get $c'^{(l)}$, because he can get $K_A$ by adding it to the receiving string $K_{A}\oplus c^{(l)}$. However, using $e^{(l)},e^{(l-1)}...,e^{(1)}$ to reconstruct $e$ is too complicated to be finished generally. Thus he has to do the step 4 to get the $c'^{(l)}$, and then to get the $K_A'$. Thus we can see that the key redistribution protocol is more suitable than the ECC based IR protocol for being reconstructed into a concatenated form.

\section{Discussions}
Concatenated IR scheme can reduce the error rate to any given level if and only if every error correction round makes the error rate lower. Thus, if the error rate of the channel satisfies Eq.(8), after a few error correction round, we can arrive at an error rate less than the given value. We choose the complete Hamming code $[2^k-1, 2^{k-1}-1-k, 3]$ to do this because of their rapid decoding algorithm. The result shows that the error rate decreases exponentially with the concatenated depth.

Error rate estimation via public channel is another basic step of QKD. It is usually an interactive process. We can leave it out by using concatenating IR scheme. For a given error rate of the raw key, after the first round syndrome calculating, the rate of non-zero syndromes should be less than a threshold. e.g., if the given error rate is $p$, the non-zero rate of syndromes of the first error correction round is less than $(1-p)^n$. If the rate is beyond this threshold, Bob simply informs Alice to give up this packet. Otherwise, Bob continues his process. In QKD, after the base sifting step, the classical data post-processing, together with error estimation using our method, can be constructed into a single protocol with almost one-way classical communication.

We can see that there are at least three interactions in a BB84 QKD protocol. The first one is quantum signal transmission from Alice to Bob; The second one is measurement information transmission from Bob to Alice: Bob informing Alice the positions of qubits received and the bases of his measurement; The third one is a classical packet from Alice to Bob: a bit string representing the positions of raw key bits she selected, and a sequence of syndromes, Alice puts them in a packet and sends it to Bob. Then Bob does the error rate check and the post-processing described above. If Bob finds the non-zero rate of syndrome is bigger than $(1-p)^{n}$, he has to do the fourth interaction to inform Alice abandoning that packet.

The concatenated IR method cannot reduce the information leakage rate. Because the adversary cannot predict the positions of his eavesdropped bits in the raw key, the eavesdropped bits are uniformly located in both the information digits and the check digits of the raw key's codewords. After each error correction round, the left bit string is permuted by wire link permutation. Thus the left leaking bits will be uniformly distributed in both the information digits and the check digits of the next round's blocks. Suppose the eavesdropping rate of the adversary is $\eta$. After abandoning the check bits in each error correction round, the length of the block is decreased from $n$ bits to $k$ bits. After $l$ rounds error correction, there are $(\frac{k}{n})^l\eta n$ bits information leakage left. Thus, after $l$ rounds reconciliation, the final information leakage rate is still $\eta$, and the parameters of privacy amplification remains the same.

\section{Conclusion}
we suggest a concatenating way to improve the efficiency of IR schemes, and construct three one-way concatenated IR schemes for QKD. The IR schemes designed based on this idea can work with only one time one-way communication and achieve any given error rate level, thus may improve the efficiency of a QKD's post-processing. In addition, a QKD scheme with this kind of IR may omit a special interaction of error rate estimation.

\section*{Acknowledgement}
This work was supported by National Natural Science Foundation of China under Grant No. 61173157.
\appendix

\section{The derivation of Eq.(8)}
\begin{eqnarray}
C^k_n-A_k & = &
     C^k_n-\frac{1}{n+1}C^k_n-\frac{n}{n+1}(-1)^{\lceil \frac{k}{2}\rceil}C^{\lfloor \frac{k}{2}\rfloor}_{\frac{n-1}{2}}\nonumber\\
     & = &
     \frac{n}{n+1}(C^k_n-(-1)^{\lceil \frac{k}{2}\rceil}C^{\lfloor \frac{k}{2}\rfloor}_{\frac{n-1}{2}}).
\end{eqnarray}
\begin{eqnarray}
& &
\frac{A_{k+1}-A_{k-1}}{A_{k-1}+A_{k+1}} \nonumber\\
     & = &
     \frac{\frac{1}{n+1}C^{k+1}_n+\frac{n}{n+1}(-1)^{\lceil \frac{k+1}{2}\rceil}C^{\lfloor \frac{k+1}{2}\rfloor}_{\frac{n-1}{2}}-\frac{1}{n+1}C^{k-1}_n-\frac{n}{n+1}(-1)^{\lceil \frac{k-1}{2}\rceil}C^{\lfloor \frac{k-1}{2}\rfloor}_{\frac{n-1}{2}}}{\frac{1}{n+1}C^{k+1}_n+\frac{n}{n+1}(-1)^{\lceil \frac{k+1}{2}\rceil}C^{\lfloor \frac{k+1}{2}\rfloor}_{\frac{n-1}{2}}+\frac{1}{n+1}C^{k-1}_n+\frac{n}{n+1}(-1)^{\lceil \frac{k-1}{2}\rceil}C^{\lfloor \frac{k-1}{2}\rfloor}_{\frac{n-1}{2}}}\nonumber\\
     & = &
     \frac{C^{k+1}_n-C^{k-1}_n+n(-1)^{\lceil \frac{k+1}{2}\rceil}(C^{\lfloor \frac{k+1}{2}\rfloor}_{\frac{n-1}{2}}+C^{\lfloor \frac{k-1}{2}\rfloor}_{\frac{n-1}{2}})}{C^{k+1}_n+C^{k-1}_n+n(-1)^{\lceil \frac{k+1}{2}\rceil}(C^{\lfloor \frac{k+1}{2}\rfloor}_{\frac{n-1}{2}}-C^{\lfloor \frac{k-1}{2}\rfloor}_{\frac{n-1}{2}})}\nonumber\\
     & = &
     \frac{A+B}{C+D},
\end{eqnarray}
Here,
\begin{eqnarray}
     A
     & = &
     (n-1)!(\frac{1}{(k+1)!(n-k-1)!}-\frac{1}{(k-1)!(n-k+1)!})\nonumber\\
     & = &
     C^{k+1}_{n-1}\frac{n^2+n-4k}{(n-k-1)(n-k)(n-k+1)},
\end{eqnarray}
\begin{eqnarray}
     B
     & = &
     (-1)^{\lceil \frac{k+1}{2}\rceil}(\frac{n-1}{2})!\frac{1}{\lfloor \frac{k+1}{2}\rfloor!(\frac{n-1}{2}-\lfloor \frac{k-1}{2}\rfloor)!}(\frac{n-1}{2})\nonumber\\
     & = &
     (-1)^{\lceil \frac{k+1}{2}\rceil}C^{\lfloor \frac{k+1}{2}\rfloor}_{\frac{n+1}{2}},
\end{eqnarray}
\begin{eqnarray}
     C
     & = &
     (n-1)!(\frac{1}{(k+1)!(n-k-1)!}+\frac{1}{(k-1)!(n-k+1)!})\nonumber\\
     & = &
     C^{k+1}_{n-1}\frac{n^2+n+2k^2-2k}{(n-k-1)(n-k)(n-k+1)},
\end{eqnarray}
\begin{eqnarray}
D & = &
     (-1)^{\lceil \frac{k+1}{2}\rceil}(\frac{n-1}{2})!\frac{\frac{n+1}{2}-\lfloor \frac{k+1}{2}\rfloor-\lfloor \frac{k+1}{2}\rfloor}{\lfloor \frac{k+1}{2}\rfloor!(\frac{n+1}{2}-\lfloor \frac{k+1}{2}\rfloor)!}\nonumber\\
     & = &
     (-1)^{\lceil \frac{k+1}{2}\rceil}(C^{\lfloor\frac{k+1}{2}\rfloor}_{\frac{n+1}{2}}-2C^{\lfloor\frac{k-1}{2}\rfloor}_{\frac{n-1}{2}})\nonumber\\
     & = &
     (-1)^{\lceil \frac{k+1}{2}\rceil}C^{\lfloor\frac{k+1}{2}\rfloor}_{\frac{n+1}{2}}(1-\frac{4}{n+1}\lfloor\frac{k+1}{2}\rfloor).
\end{eqnarray}
Thus,
\begin{eqnarray}
\frac{A_{k+1}-A_{k-1}}{A_{k-1}+A_{k+1}}
&\!\!\! =\!\! &
\frac{C^{k+1}_n-C^{k-1}_n+n(-1)^{\lceil \frac{k+1}{2}\rceil}C^{\lfloor \frac{k+1}{2}\rfloor}_{\frac{n+1}{2}}}{C^{k+1}_n+C^{k-1}_n+n(-1)^{\lceil \frac{k+1}{2}\rceil}C^{\lfloor \frac{k+1}{2}\rfloor}_{\frac{n+1}{2}}(1-\frac{4}{n+1}\lfloor\frac{k+1}{2}\rfloor)},
\end{eqnarray}
Here,
\begin{equation}
    C^{k+1}_n+C^{k-1}_n=C^{k+1}_{n+1}\frac{n^2+n-2nk+2k^2}{n^2+2n+1-k},
\end{equation}
\begin{eqnarray}
C^{k+1}_n-C^{k-1}_n
& = &
\frac{n!}{(k+1)!(n-k-1)!}-\frac{n!}{(k-1)!(n-k+1)!}\nonumber\\
& = &
\frac{n!}{(k+1)!(n-k+1)!}[(n-k+1)(n-k)-(k+1)k]\nonumber\\
& = &
C^{k+1}_{n+1}\frac{n-2k}{n-k+1}.
\end{eqnarray}

\section{The proof of Lemma 1}
\textbf{Proof.} Hamming code can correct one-bit error without failure. When there are more errors, the correction process may add 1 bit error. Here we consider the upper bound of the average number of errors, thus we assume the number of errors will increase by 1 after error correcting. When there are $n$ bits errors, the number of errors will be reduced by 1 after error correction. Then

\begin{equation}
\begin{split}
\chi & = \sum_{k=2}^{n-1}(1+k)C_n^kp^k(1-p)^{n-k}+(n-1)C_n^np^n \\
     & = \sum_{k=2}^{n}(1+k)C_n^kp^k(1-p)^{n-k}-2p^n \\
     & = \sum_{k=0}^{n}(1+k)C_n^kp^k(1-p)^{n-k}-2p^n-(1-p)^n-2np(1-p)^{n-1}.
\end{split}
\end{equation}

\noindent By the identity $\sum_{k=0}^{n}kC_n^kp^k(1-p)^{n-k}=np$, we have

\begin{equation}
\chi = 1+np-2p^n-(1-p)^{n-1}(1-p+2np).
\end{equation}
$\hfill{}\Box$

Now let us consider the upper bound of $\chi$.

\section{The proof of Lemma 2}
\textbf{Proof:} From the Eq.(B.2), we have
\begin{equation}
\begin{split}
\chi & < \sum_{k=2}^{n-1}(1+k)C_n^kp^k(1-p)^{n-k} \\
     & = \sum_{k=3}^{n}(1+k)C_n^kp^k(1-p)^{n-k}+3C_n^2p^2(1-p)^{n-2}.
\end{split}
\end{equation}

\noindent By the inequality\cite{16} $(1+k)C_n^k\leq n(n-1)C_{n-2}^{k-2}$ $(k\geq 3)$, it holds that
\begin{equation}
\begin{split}
\sum_{k=3}^{n}(1+k)C_n^kp^k(1-p)^{n-k} & \leq n(n-1)\sum_{k=3}^{n}C_{n-2}^{k-2}p^k(1-p)^{n-k} \\
                                       & = n(n-1)p^2\sum_{k=3}^{n}C_{n-2}^{k-2}p^{k-2}(1-p)^{n-k} \\
                                       & = n(n-1)p^2\sum_{k=1}^{n-2}C_{n-2}^{k}p^{k}(1-p)^{n-2-k} \\
                                       & = n(n-1)p^2[1-(1-p)^{n-2}].
\end{split}
\end{equation}

\noindent Thus we obtain
\begin{equation}
\begin{split}
\chi & < 2C_n^2p^2[1-(1-p)^{n-2}]+3C_n^2p^2(1-p)^{n-2} \\
     & = n(n-1)p^2[1+\frac{1}{2}(1-p)^{n-2}].
\end{split}
\end{equation}
$\hfill{}\Box$

\noindent From the Lemma 2, it holds that
\begin{equation}
\chi < \frac{3n(n-1)}{2}p^2 < \frac{3}{2}(np)^2.
\end{equation}

\section{The proof of Theorem 1}
\textbf{Proof.} Denote $p_1$ as the error rate after one error correction round. From the definition of $\chi$, we know $p_1<\frac{\chi}{n}.$ It is clear that the concatenated error correction scheme can reduce the error rate to any given level, if and only if $p_1<p$. Because $p_1<\frac{\chi}{n}$, $p_1<p$ holds if $\frac{\chi}{n}<p$. From Lemma 2, $\frac{\chi}{n}<p$ holds if $n(n-1)p^2[1+\frac{1}{2}(1-p)^{n-2}]<np$. That is
\begin{equation}
p<\frac{1}{(n-1)[1+\frac{1}{2}(1-p)^{n-2}]}.
\end{equation}
$\hfill{}\Box$

\section{The proof of Corollary 1}
\begin{equation}
\frac{2}{3(n-1)}<\frac{1}{(n-1)[1+\frac{1}{2}(1-p)^{n-2}]}<\frac{1}{n-1}.
\end{equation}
Thus, when $p<\frac{2}{3(n-1)}$, the condition Eq.(D.1) holds. Let $p_{th}=\frac{2}{3(n-1)}$. Thus if $p<p_{th}$, according to Theorem 1, the concatenated error correction scheme can reduce the error rate to any given level.
$\hfill{}\Box$

%% References


%\bibliographystyle{model1a-num-names}



\end{document}